# The sign of the polarizability anisotropy of polar molecules is obtained faithfully from terahertz Kerr effect

**Tobias Kampfrath\*, Martin Wolf and Mohsen Sajadi\***
*Fritz-Haber-Institut der Max-Planck-Gesellschaft, Berlin, Germany*

**\***kampfrath@fhi-berlin.mpg.de, sajadi@fhi-berlin.mpg.de.

**Optically heterodyned detected terahertz Kerr effect of gases of polar molecules is reported. Strikingly, the birefringence signal from fluoroform is found to have opposite polarity compared to water and acetonitrile. This behavior is a hallmark of the opposite sign of the polarizability anisotropy $\Delta\alpha$ of these molecules. Resonant excitation of the rotational degrees of freedom of the molecules aligns their permanent dipoles along the terahertz electric field. This alignment is translated into an optical birefringence through $\Delta\alpha$ of each molecule. Therefore, the resulting net signal scales with $\Delta\alpha$, whose sign is imprinted faithfully onto the transient birefringence signal.**

Understanding the structural dynamics of complex molecular systems requires insights into the underlying intermolecular interactions.[1,2,3,4,5,6,7] To address this goal, one may employ spectroscopic techniques ranging from time-averaged dielectric spectroscopy[8] to time-resolved multidimensional spectroscopies,[2] in conjunction with molecular-dynamics simulations.[9] Here, recent advances in nonlinear terahertz (THz) spectroscopy offer new opportunities as intense THz electromagnetic waves can interrogate and even drive low-frequency molecular resonances directly.[10,11,12,13,14,15]

For example, rotational modes are often located in the THz regime. In the case of polar molecules, the THz electric field $E(t)$ can exert a torque by direct coupling to the permanent molecular dipole moment $\mu_0$, thereby aligning the molecular dipoles along the direction of $E$. The associated interaction energy is given by $V_{\mathrm{perm}}(\theta,t) = -\mu_0 E(t)\cos\theta$ where $\theta$ is the angle between the directions of the THz field and the permanent molecular dipole.[10] This mechanism, however, is not operative for optical light fields because the molecules cannot follow the rapid field oscillations.[16]

On the other hand, molecular alignment can also be achieved by coupling to the electronic dipole moment that is transiently induced by the applied electric field. For molecules with cylindrical symmetry, the coupling energy is given by $V_{\mathrm{ind}}(\theta,t) = -(\Delta\alpha/2)E(t)^2\cos^2\theta$, where $\Delta\alpha = \alpha_\parallel - \alpha_\perp$ is the difference between the electronic polarizabilities parallel and perpendicular to the symmetry axes of the molecules. The time-averaged torque on the molecules does not vanish because the squared electric field in $V_{\mathrm{ind}}$ results in rectification of the rapidly varying light field.[16] As a result, molecules will be aligned with the axis of their largest polarizability parallel to the driving field polarization.

Note that the $V_{\mathrm{perm}}$-coupling aligns the molecular dipoles along the direction of the driving field, whereas for $V_{\mathrm{ind}}$, the relevant axis is the one with the largest molecular polarizability. Therefore, two different angular distributions may be formed after an optical and a THz excitation, respectively. As in general, the permanent dipole moment $\boldsymbol{\mu}_0$ of a molecule is easier to determine than its polarizability tensor elements, an aligned molecular ensemble via $V_{\mathrm{perm}}$ interaction is easier to characterize than the one aligned with $V_{\mathrm{ind}}$ interaction, too. A unique property of permanent-dipole-type alignment is that the optical polarizability of molecules remains unchanged and, thus, can be accessed by a suitable probe.

The instantaneous amplitude of the molecular alignment and its relaxation dynamics can be measured by detecting the refractive-index difference $\Delta n$ between the directions parallel and perpendicular to the driving linearly polarized THz field (THz Kerr effect, TKE).[17] The strength of the transient optical birefringence can be measured by an optical probe pulse and is given by [14]

.

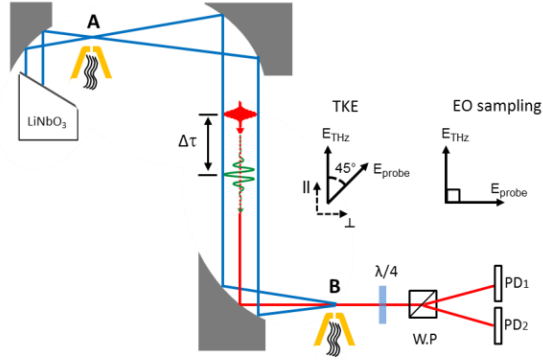

**Fig. 1. Schematic of the THz setup.** To measure THz-induced optical birefringence of gases, molecules of the sample gas are brought into the THz focus at position **B** by means of a nozzle. At this position, the transient birefringence induced by the THz pump pulse is measured by the collinearly propagating optical probe. For THz transmission measurements, the nozzle is placed at position **A**. After having traversed the sample gas, the transient THz electric field is sampled in an electrooptic (110)-oriented ZnTe crystal at position **B**. Abbreviations: λ/4: quarter waveplate, W.P: Wollaston prism, PD: photodiode.

$$\Delta n(t) \propto \Delta\alpha \int d\cos\theta\ f(\theta,t)(\cos^2\theta - 1/3) \quad (1)$$

where $f(\theta,t)$ is the molecular rotational distribution function at time $t$ after THz excitation.

A schematic of our experiment is shown by **Fig. 1**. A gas jet of the sample molecules at position B is excited by intense, linearly polarized THz pulses from a Lithium Niobate (LN) source with field strength exceeding 2 MV cm$^{-1}$. The resulting transient optical birefringence is sampled by a time-delayed optical pulse (pulse energy 2 nJ, duration 8 fs, center wavelength 800 nm) whose linear polarization before the sample is at 45° with respect to the direction of the THz field. After propagation through the gas, a phase difference between the probe field components parallel and perpendicular to the THz field has accumulated, which is proportional to $\Delta n$. It is measured by heterodyne detection using a quarter-wave plate and a balanced optical bridge (**Fig. 1**). The resulting THz amplitude spectrum is shown in **Fig. 2a** (yellow area). For all experiments, the setup is purged with N$_2$. More details of the setup and the THz source are given elsewhere.[18]

As gas targets, we use vapors of acetonitrile (CH$_3$CN), water (H$_2$O) and fluoroform (CHF$_3$), as the molecules are all polar ($\mu_0 = 3.9$, 1.8 and 1.65 D, respectively) but exhibit polarizability anisotropies of different sign ($\Delta\alpha = 2.15$, 0.098 and $-0.18$ Å$^3$, respectively). Acetonitrile (99.9% from Sigma-Aldrich) and water (double-distilled) are kept inside a small vessel and heated up to 55 °C and 75 °C, respectively,[19,20,21] while the fluoroform (from Air Liquide)[22,23] is already gaseous. The gas molecules are brought to the THz focus via a nozzle which has a hole with a diameter of 0.5 mm. (see **Fig. 1**). The distance between the liquid surface and the THz focuses is about 5 mm. We locate a sucking tube on top of the THz focus and the nozzle to send the gases out of the purging box. We estimate the thickness of the interaction region to be about 1 mm.

To ensure reliable comparison of the various TKE signals, we performed all experiments within one day and under the same experimental conditions. In particular, the polarization states of the THz pump and optical probe pulses, as well as the alignment of quarter waveplate are identical for all measurements.

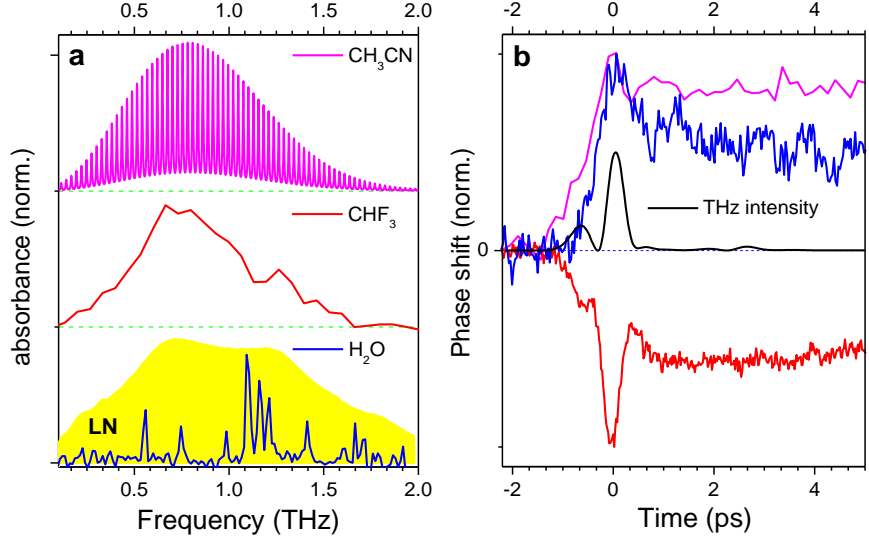

**Fig. 2**. **Linear and nonlinear THz results**. (**a**) Red and blue solid lines show measured THz absorption spectra of fluoroform and water vapor, respectively. The solid magenta line shows the simulated THz absorption spectrum of acetonitrile vapor[26, 27]. The yellow area is the amplitude spectrum of the THz pump pulse from the LN source. (**b**) Birefringence signals of acetonitrile (magenta line) and water (blue line) vapors and of fluoroform (red line). The instantaneous THz intensity $E^2(t)$ is shown by the black line.

To obtain linear THz absorption spectra of water and fluoroform, the nozzle is placed at position **A** and a ZnTe(110) electrooptic crystal at position **B**. The THz power is lowered to avoid detection saturation. For the electrooptic sampling, the THz pump and the 800 nm probe pulses have a relative polarization angle of 90°. The absorption spectra of the water vapor and the fluoroform are calculated as $-2\ln|E(\omega)/E_0(\omega)|$, where $E(\omega)$ and $E_0(\omega)$ are the THz electric field amplitudes at frequency $\omega/2\pi$ with and without sample gas, respectively.

The measured THz absorption spectra of fluoroform and water vapor are shown in **Fig. 2a**. Their shape agrees well with that of previous measurements.[24,25] For water, the sharp absorption peaks are due to the rotational transitions between thermally populated states.[25] Similar peaks could not be resolved for fluoroform because of its weak echo signals. The absorption spectrum of acetonitrile vapor was not measured because of toxicity issues. It is, therefore, simulated based on the procedure given in Refs. 26 and 27 at a temperature of 328 K. The comb of sharp absorption peaks arises from rotational transitions. We emphasize that absorption of the gases over the frequency range covered by the pump is solely due to the rotational degrees of freedom of the gas molecules. Excitation of the vibrational degrees of freedom requires pump frequencies > 10 THz.[28]

The transient birefringence signals of the gas targets following THz excitation at time zero up to delays of 5 ps are shown in **Fig. 2b**. All signals are unipolar and exhibit a step-like increase which occurs within the duration of the THz pump pulses (see black solid line in **Fig. 2b**). All signals decay back to zero on a time scale of 100s ps. We also find sharp temporal features (revivals) after time intervals on the order of 20-30 ps for acetonitrile and fluoroform (data not shown).

Strikingly, the signals for acetonitrile and water are found to be positive, whereas a negative signal is observed for fluoroform. In the following discussion, we focus on the nature of this sign difference and identify a molecular parameter by which we can explain the sign change of the TKE signal.

In the frequency range covered by our THz pump, dynamics are dominated by rotational degrees of freedom (see above). In general, the transient birefringence signals in the gas phase consist of coherent and non-coherent (population) contributions to molecular alignment.[10] Coherent effects appear as sharp temporal features (revivals) in the birefringence signals with time intervals dictated by the fundamental rotational constants of the molecules. Here, we are interested in the non-coherent portion of the signals that are also due to the alignment of the molecules.[29] Recently, we showed that for molecular ensembles with only rotational degrees of freedom, an extended version of the rotational-diffusion model provides a basic understanding of the alignment process by THz and optical fields.[14] When a system of static rotors is excited with an electromagnetic wave, the induced ensemble-averaged alignment of molecules can be expressed as

$$\langle \cos^2 \theta(t) - 1/3 \rangle \propto R_2 * [E \cdot \{N\Delta\alpha E + 3\chi^{\mu_0} * E\}]. \qquad (2)$$

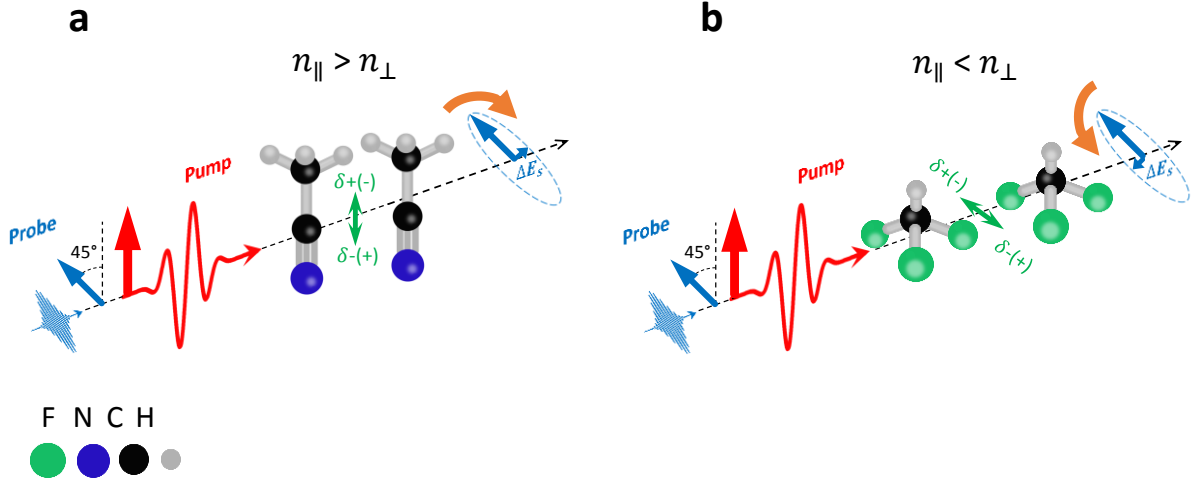

**Fig. 3**. **Picture of the THz alignment of dipolar molecules with opposite $\Delta\alpha$ sign**. (**a**) After interaction of the THz field with the permanent molecular dipole moments, the symmetry axis of both the acetonitrile ($\Delta\alpha > 0$) and (**b**) the fluoroform ($\Delta\alpha < 0$) molecule is aligned along the direction of the THz electric field. The resulting molecular alignment is probed by an incident linearly polarized laser pulse which acquires ellipticity after propagating through the ensemble of aligned molecules. The probe field polarizes the electron cloud of the molecules in a direction given by $\Delta\alpha$, shown by double-side arrows. Accordingly, the probe pulse experiences opposite birefringence in acetonitrile and fluoroform and, thus, acquires ellipticity with opposite handedness, thereby resulting in a sign change of the TKE signal.

Here, $N$ is the number of molecules per volume, the function $R_2(t)$ describes the decay of the resulting anisotropy in the molecular ensemble, $N\Delta\alpha E$ is the electronic polarization, and $\chi^{\mu_0}$ stands for the contribution of the $V_{\text{perm}}$-related torque to the dielectric susceptibility.

The first term of Eq. (2) is due to the coupling $V_{\text{ind}}$ of the pump field to the induced electronic dipole whereas the second term arises from the coupling $V_{\text{perm}}$ to the permanent molecular dipole moment. Note that Eq. (2) can be understood as a two-step process. In the first step, the incident electric field induces a polarization in the medium with two different contributions arising from the electronic polarization ($N\Delta\alpha E$) and orientational polarization ($3\chi^{\mu_0} * E$), respectively. The second field interaction turns the polarization ($\propto \langle \cos \theta(t) \rangle$) into the alignment ($\propto \langle \cos^2 \theta(t) - 1/3 \rangle$).

As the optical probe pulse measures the alignment by coupling through $\Delta\alpha$ (see Eq. (1)), the birefringence signal scales with $\Delta\alpha^2$ when alignment is due to the torque exerted through the induced electronic dipole moment (see Eq. (2)). Therefore, from an optical Kerr effect (OKE) experiment, no information on the sign of $\Delta\alpha$ can be obtained.

In contrast, for THz pumping and the polar gases considered here, one has $\chi^{\mu_0} * E \gg N\Delta\alpha E$.[30] In other words, the TKE signal is dominated by coupling of the THz field to the permanent molecular dipole moments and, thus, proportional to $\Delta\alpha$. We conclude that the sign of the birefringence traces of **Fig. 2b** is a clear hallmark of the sign of the polarizability anisotropy of the molecules studied: positive for acetonitrile and water yet negative for fluoroform.

This finding can be illustrated by the picture suggested in **Fig. 3**: In the case of acetonitrile and water, the direction of the largest polarizability and the permanent dipole moment are along the symmetry axis of the molecule, implying a positive $\Delta\alpha$. For fluoroform, in contrast, the polarizability of the CF bonds, $\alpha_{xx}$ =2.898 Å$^3$ is larger than the polarizability component of the molecule along its dipole (symmetry) axis $\alpha_{zz}$ = 2.718 Å$^3$ (**Fig. 3**), thereby resulting in a negative $\Delta\alpha$.[23] In the case of perfect alignment by the THz field, the permanent dipole moments of both molecules point in the THz field direction (**Fig. 3**). However, as acetonitrile and fluoroform have opposite $\Delta\alpha$ sign, their slow optical axes (i.e. the axes with largest polarizability) are approximately parallel and perpendicular to their symmetry axes, respectively. Accordingly, acetonitrile and fluoroform acquire positive and negative birefringences relative to the polarization axis of the THz pump. As a result, a linearly polarized probe pulse traversing an ensemble of either of the gases acquires ellipticity with opposite handedness. The latter handedness can be resolved in an optically heterodyned detection configuration as it scales linearly with $\Delta n$ (see Eq. (1)).

In conclusion, we showed that the sign of the polarizability anisotropy of dipolar molecules can be obtained directly from the THz Kerr effect. In fact, the combination of two features, namely (i) the coupling of the THz fields to the permanent molecular dipole moments and (ii) the heterodyne detection enables us to extract the sign of $\Delta\alpha$ for polar molecules. Our results are also consistent with previous works reporting that $\Delta\alpha$ of water vapor is positive.[31] Interestingly, the sign of $\Delta\alpha$ cannot faithfully be obtained from experiments based on the optical Kerr effect[32] or depolarized spontaneous Raman scattering[33]. This goal requires additional information, e.g. obtained by "gaining experience",[34] molecular simulations or theory.[22,35]

**References**

(1) Berne, B. J.; Pecora, R. Dynamic light scattering with applications to chemistry, Biology and Physics (Dover, New York, 2000).
(2) Mukamel, S. Principles of nonlinear optical spectroscopy, (Oxford University Press, New York, 1995).
(3) Maroncelli, M. The dynamics of solvation in polar liquids. *J. Molec. Liq.* **1993** 57, 1-37.
(4) Birnbaum, G. Phenomena induced by intermolecular interactions (Plenum, New York, 1985)
(5) Lamley, J. M.; Öster, C.; Stevens, R. A.; Lewandowski, J. R. Intermolecular Interactions and Protein Dynamics by Solid-State NMR Spectroscopy. *Angew. Chem. Int. Ed.* **2015**, 54, 15374–15378.
(6) Zusman, L. D. Outer-sphere electron-transfer in polar-solvents. *Chem. Phys.* **1980**, 49, 295-304.
7 Tielrooij, K. J.; Garcia-Araez, N.; Bonn, M.; Bakker, H. J. Cooperativity in Ion Hydration. *Science,* **2010**, 328, 1006-1009.
(8) Kremer, F.; Schönhals, A. Broadband Dielectric Spectroscopy (Berlin: Springer, 2003)
(9) Rapaport, D. C. The Art of Molecular Dynamics Simulation (Cambridge University Press, Cambridge, England, 1998)
(10) Fleischer, S.; Zhou, Y.; Field, R. W.; Nelson, K. A. Molecular orientation and alignment by intense single-cycle THz pulses. Phys. Rev. Lett. **2011**, 107, 163603–163605.
(11) Allodi, M. A.; Finneran, I. A.; Blake, G. A.; Nonlinear terahertz coherent excitation of vibrational modes of liquids. J. Chem. Phys. **2015**, 143, 234204–234208.
(12) Finneran, I. A.; Welsch, R.; Allodi, M. A.; Miller, T. F.; Blake, G. A. Coherent two-dimensional terahertz-terahertz-Raman spectroscopy. *Proc. Natl. Acad. Sci. USA*. **2016**, 113, 6857–6861.
(13) Hoffmann, M. C.; Brandt, N. C.; Hwang, H. Y.; Yeh, K. L.; Nelson, K. A. Terahertz Kerr effect. *Appl. Phys. Lett.* **2009**, 95, 231105.
(14) Sajadi, M.; Wolf, M.; Kampfrath, T. Transient birefringence of liquids induced by terahertz electric-field torque on permanent molecular dipoles. *Nat. Commun.* **2017**, 8, 14963.

(15) Lua, J.; Zhanga, Y.; Hwanga, H. Y.; Ofori-Okaia, B. K.; Fleischer, S.; Nelson, K. A. Nonlinear two-dimensional terahertz photon echo and rotational spectroscopy in the gas phase. *Proc. Natl. Acad. Sci. USA*. **2016**, 113, 11800–11805.

(16) Stapelfeldt, H.; Seideman, T.; Aligning molecules with strong laser pulses. Rev. Mod. Phys. **2003**, 75, 543-557

(17) Righini, R. Ultrafast optical Kerr-effect in liquids and solids. Science, **1993**, 262, 1386–1390.

(18) Sajadi, M., Wolf, M. & Kampfrath, T. Terahertz-field-induced optical birefringence in common window and substrate materials. *Opt. Express*, **2015**, 23, 28985–28992.

(19) Riddick, J. A.; Bunger, W.B.; Sakano, T. K. Organic Solvents (Wiley: New York, 1986).

(20) Suttert, H.; Cole, R. H. Dielectric and Pressure Virial Coefficients of Imperfect Gases. I. Polar Halogenated Methanes, *J. Chem. Phys*. **1970**, 52, 132-139.

(21) Nir, S.; Adams, S.; Rein, R. Polarizability calculations on water, hydrogen, oxygen, and carbon dioxide. *J. Chem. Phys*. **1973**, 59, 3341-3355.

(22) Miller, C. K.; B. J. Orr, B. J.; Ward, J. F. An interacting segment model of molecular electric tensor properties: Application to dipole moments, polarizabilities, and hyperpolarizabilities for the halogenated methanes. *J. Chem. Phys*., **1981**, 74, 4858-4871.

(23) Kobayashi, R.; Amos, R. D.; Koch, H.; Jørgensen, P. Dynamic CCSD polarisabilities of $CHF_3$ and $CHCl_3$, *Chem. Phys. Lett*. **1996**, 253, 373-376.

(24) Saitow, K.; Ohtake, H.; Sarukura, N.; Nishikawa, K. Terahertz absorption spectra of supercritical CHF3 to investigate local structure through rotational and hindered rotational motions, *Chem. Phys. Lett.* **2001**, 341, 86-92.

(25) van Exter, M.; Fattinger, C. Grischkowsky, D. Terahertz time-domain spectroscopy of water vapor, *Opt. Lett*. **1989**, 14, 1128-1130.

(26) Harde, H.; Katzenellenbogen, N.; Grischkowsky, D. Terahertz coherent transients from methyl chloride vapor, *J. Opt. Soc. Am. B*, **1994**, 11, 1018-1030.

(27) Melinger, J. S.; Yang, Y.; Mandehgar, M.; Grischkowsky, D.; THz detection of small molecule vapors in the atmospheric transmission windows, Opt. Express, **2012**, 20, 6788- 6807.

(28 ) Townes, C. H.; Schawlow, A. L. Microwave Spectroscopy, (New York, McGraw-Hill, 1955).

(29) Ramakrishna S. Seideman, T. Intense Laser Alignment in Dissipative Media as a Route to Solvent Dynamics, *Phys. Rev. Lett*. **2005**, 95, 113001.

(30) For water, acetonitrile and fluoroform $\chi^{\mu_0} * E/N\Delta\alpha E \propto \frac{\beta\mu_0^2}{3\Delta\alpha} \approx 300$, 60 and 10, respectively, where $\beta = K_B T$.

(31) Murphy, W. F. The Rayleigh depolarization ratio and rotational Raman spectrum of water vapor and the polarizability components for the water molecule, *J. Chem. Phys*. **1977**, 67(12), 5877-5882

(32) Bogaard, M. P.; Orr, B. J. Electric Dipole Polarizabilities of Atoms and Molecules, in *International Review of Science, Physical Chemistry*, Series Two, Vol. 2, edited by A. D. Buckingham (Butterworths, London, 1975), pp. 149-194.

(33) Alms, G. R.; Burnham, A. K.; Flygare, W. H. Measurement of the dispersion in polarizability anisotropies, The Journal of Chemical Physics, **1975**, 63, 3321-3326.

(34) Bogaard, M. P.; Buckingham, A. D.; Pierens, R. K.; White, A. H. Rayleigh Scattering Depolarization Ratio and Molecular Polarizability Anisotropy for Gases. *J. Chem. Soc., Faraday Trans*. **1978**, 74 3008-3015.

(35) Huiszoon, C. Ab initio calculations of multipole moments, polarizabilities and isotropic long range interaction coefficients for dimethylether, methanol, methane, and water. *Mol. Phys*. **1986**, *58*, 865-886.